\documentclass[aps,prd,reprint,preprintnumbers,onecolumn]{revtex4-2}
\usepackage{amsmath, mathrsfs, amssymb,amsfonts,amsthm,graphicx, epsf, dcolumn, yfonts}
\usepackage[hyperfootnotes=true]{hyperref}

\usepackage{amsmath,epsfig}
\usepackage{amssymb,amsfonts}
\usepackage{latexsym}
\usepackage[latin1]{inputenc}
\usepackage{overpic}
\usepackage{subeqnarray}
\usepackage{array}
\usepackage{xcolor}

\usepackage{graphicx}
\usepackage{subfigure}
\usepackage{longtable}
\usepackage{mathdots}
\usepackage{physics}
\usepackage{enumitem}
\usepackage{oubraces}
\usepackage{cancel}

\relax

\def\be{\begin{equation}}
\def\ee{\end{equation}}
\def\bea{\begin{eqnarray}}
\def\eea{\end{eqnarray}}
\def\bs{\begin{subequations}}
\def\es{\end{subequations}}

\newcommand\fverb{\setbox\pippobox=\hbox\bgroup\verb}
\newcommand\fverbdo{\egroup\medskip\noindent%
                        \fbox{\unhbox\pippobox}\ }
\newcommand\fverbit{\egroup\item[\fbox{\unhbox\pippobox}]}

\newcommand{\bear}{\begin{eqnarray}}
\newcommand{\eear}{\end{eqnarray}}

\newcommand{\bsea}{\begin{subeqnarray}}
\newcommand{\esea}{\end{subeqnarray}}
\newbox\pippobox

\def\lab{\label}

\let\oldsqrt\sqrt
\def\sqrt{\mathpalette\DHLhksqrt}
\def\DHLhksqrt#1#2{%
\setbox0=\hbox{$#1\oldsqrt{#2\,}$}\dimen0=\ht0
\advance\dimen0-0.2\ht0
\setbox2=\hbox{\vrule height\ht0 depth -\dimen0}%
{\box0\lower0.4pt\box2}}

\def\lab{\label}
\def\6{\partial}

\def\a{\alpha}

\def\le{\left}
\def\ri{\right}

\def\sq
\def\a{\alpha}

\def\<{{\langle}}
\def\>{{\rangle}}
\def\le{\left}
\def\ri{\right}
\def\dg{\dagger}

\def\6{\partial}

\begin{document}

\preprint{IFT-UAM/CSIC-23-79}

\title{Universal rapidity scaling of entanglement entropy inside hadrons\\ from conformal invariance}

\author{Umut G\"ursoy,$^{1}$ Dmitri E. Kharzeev,$^{2,3}$ and Juan F. Pedraza$^{4}$}
\affiliation{\vspace{1mm}$^1$Institute for Theoretical Physics and Center for Extreme Matter and Emergent Phenomena, Utrecht University, Leuvenlaan 4, 3584 CE Utrecht, The Netherlands\\
$^2$Center for Nuclear Theory, Department of Physics and Astronomy,
Stony Brook University, Stony Brook, New York 11794-3800, USA\\
$^3$Department of Physics, Brookhaven National Laboratory, Upton, New York 11973-5000, USA\\
$^4$Instituto de F\'isica Te\'orica UAM/CSIC, Calle Nicol\'as Cabrera 13-15, Madrid 28049, Spain}

\begin{abstract}\vspace{-2mm}
\noindent
When a hadron is probed at high energy, a nontrivial quantum entanglement entropy inside the hadron emerges due to the lack of complete information about the hadron wave function extracted from this measurement. In the high-energy limit, the hadron becomes a maximally entangled state, with a linear dependence of entanglement entropy on rapidity, as has been found in a recent analysis based on parton description. In this paper, we use an effective conformal field theoretic description of hadrons on the light cone to show that the linear dependence of the entanglement entropy on rapidity found in parton description is a general consequence of approximate conformal invariance and does not depend on the assumption of weak coupling. Our result also provides further evidence for a duality between the parton and string descriptions of hadrons. 
\end{abstract}

\maketitle

\section{Introduction}
\lab{sec1}
The parton model, with its probabilistic approach to counting partons, is in apparent contradiction with quantum mechanics. Indeed, a distribution of (quasi-)free partons possesses a finite entropy, whereas the hadron should represent a pure quantum state (an eigenstate of the QCD Hamiltonian) with zero entropy. A way to resolve this paradox is to observe that in a high-energy interaction, one does not have access to entire information about the hadron state \cite{Kharzeev:2017qzs}. The density matrix of the hadron thus has to be traced over the unobserved degrees of freedom, and the resulting reduced density matrix describes a mixed state with a finite entropy. To make this proposal more concrete, consider a short pulse of light interacting with the hadron, see Fig. \ref{fig1}. The equation of the lightfront fixes $x^- = t - x = 0$, but since the positions of partons $x_i$ and moments of time $t_i$ when the light interacted with them cannot be accurately determined in a high-energy interaction, one has to trace the hadron density matrix over $x^+_i = t_i + x_i$ \cite{Kharzeev:2021nzh}. As a result, even though the overall state is pure, one gets a mixed reduced density matrix that is diagonal in the Fock basis of states with fixed numbers of partons. The entropy associated with this reduced density matrix thus results from the entanglement between the phase and parton occupation number \footnote{In the rest frame of the proton, the wave function of the virtual photon is prepared long before the interaction and does not evolve much during the interaction. We make this assumption explicit by averaging over the unobserved phase.}. 

At large rapidity (or small Bjorken $x$), the hadrons have been found to represent maximally entangled states, with the entanglement entropy growing linearly with rapidity \cite{Kharzeev:2017qzs}. This feature of the entanglement entropy inside hadrons has been confirmed in a number of approaches based on parton description \cite{Armesto:2019mna,Dvali:2021ooc,Liu:2022ohy,Liu:2022hto}. Using the relation between the entropy and the structure function proposed in \cite{Kharzeev:2017qzs}, the maximal entanglement hypothesis can be checked against the data on deep inelastic scattering \cite{H1:2020zpd}, and has been found \cite{Kharzeev:2021yyf,Hentschinski:2021aux,Hentschinski:2022rsa,Hentschinski:2023izh} to agree with experiment. Other aspects of entanglement in high-energy interactions are also under active investigation; see, e.g., \cite{Kovner:2015hga,Peschanski:2016hgk,Berges:2017hne,Berges:2018cny,Peschanski:2019yah,Florio:2023dke,miller2023entanglement}.

Because the entanglement entropy at high energies emerges from the space-time picture of the interaction, it seems natural to expect that this phenomenon should not be restricted to the parton model, and can be described in a more general setting that is  constrained only by the symmetries of the interaction. Below we will present such a description based on conformal invariance.

How reasonable is it to expect that conformal invariance is relevant in high-energy hadron interactions? In high-energy perturbative QCD, the effective action derived by Lipatov \cite{Lipatov:1993yb} possesses conformal invariance, and can be mapped onto a XXX spin chain \cite{Faddeev:1994zg} and the nonlinear Schr\"odinger equation \cite{Hao:2019cfu}. This conformal invariance of Lipatov action has been used to evaluate the entanglement entropy, which has been found to scale linearly in rapidity \cite{Zhang:2021hra} (see also \cite{Kutak:2011rb} which argued for the linear in rapidity scaling for the entropy of produced gluons). Here we will show that the linear dependence of the entanglement entropy on rapidity (corresponding to the maximally entangled state of the hadron) is a general consequence of the approximate conformal invariance on the light cone. We will briefly review how conformal invariance of hadron wave functions emerges on the light cone in the next section.

\section{Effective Conformal Field-Theoretical description of hadrons}
\lab{sec2}

\begin{figure}[t!]
\centering
\includegraphics[width=8cm]{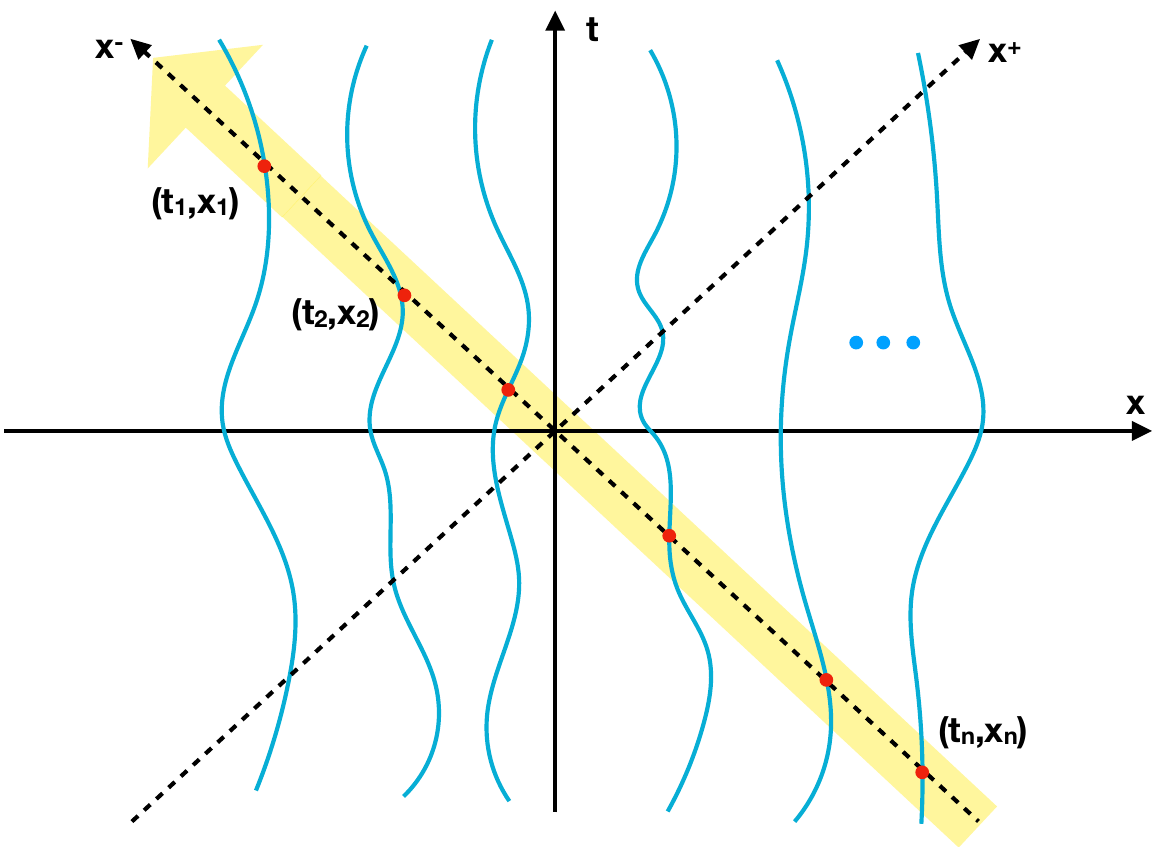}
\caption{Light front at $x^+=0$, shown by yellow, capturing positions of $n$ partons at $x_i=t_i$, shown by red dots. Full information on the worldline of partons also involves $x^+$ which the light front does not have access to, hence creating a reduced density matrix with nontrivial entanglement entropy. The partons are depicted as nearly static for illustrative purposes; however, in the context of high-energy scattering, they would be narrowly distributed around $x^-=0$.}\label{fig1}
\end{figure}

Consider a deep inelastic scattering (DIS), in which a hadron containing $n$ partons is probed by a virtual photon. Partons are created and annihilated by operators $a_i^\dg$ and $a_i$ with $i=1,\cdots, n$ and are probed on the light cone. One can alternatively describe the system in a second quantized description where the operators $a(x^+,x^-)$ are defined on the light cone.
This leads to a description of a hadron in terms of an effective quantum field theory on the 2D space-time spanned by $x^\pm$. This setup is schematically shown in Fig. \ref{fig1}.
In fact, this effective field-theoretical description applies even when the coupling constant is large. Even though we do not know the precise action of this 2D quantum field theory at strong coupling, there are good reasons to believe that it can be deduced from an effective theory of a string with a worldsheet spanning the light-cone coordinates $x^\pm$ as we explain below. 

String description of strong interactions pre-dates QCD \cite{Veneziano}. The idea of describing color-flux tubes between quark-antiquark pairs was proposed in the 1960s but it materialized as a string worldsheet theory only after the seminal works of `t Hooft \cite{thooft}, Polyakov \cite{Polyakov1}, Nambu \cite{Nambu1} and others. Today this idea has at least two incarnations: An {\em exact} string description of gauge theories as the AdS/CFT correspondence \cite{Maldacena:1997re,Witten:1998qj,Gubser:1998bc}, and a low-energy {\em effective} theory of long strings; see, for example \cite{Nambu2, LuscherWeisz1, LuscherWeisz2, Sundrum}. This latter approach is based on the fact that an open string breaks space-time symmetry as ISO(3,1)$\to$ ISO(1,1)$\times$SO(2) hence string excitations in the transverse directions are described by two Nambu-Goldstone bosons $X^I$, $I=1,2$. For example, for a meson this effective string description arises from the color-electric flux between the quark and the antiquark and is given by an action 
\be\lab{act}
S_{\text{st}} = T \int dx^+dx^- \le(  \6_+ X^I \6_- X^I + \cdots\ri)\, .
\ee
Here $T$ denotes the string tension and we fixed the worldsheet symmetries in the {\em static} gauge \cite{Aharony:2013ipa}, where worldsheet coordinates are anchored to space-time coordinates $x$ and $t$ \footnote{See \cite{Polchinski:1991ax,Arvis:1983fp} for alternative quantization schemes.}. In this gauge, the worldsheet fields, e.g. the transverse Nambu-Goto bosons (and possibly other fields) become functions of $x^\pm$ providing an effective 2D conformal field theory (CFT) in the 2D space-time spanned by $x^\pm$. Ellipsis in (\ref{act}) denote possible other terms such as $n\cdot \6 X^I A_I$, which describes the interaction of the hadron with a photon (with $n^{\pm}$ denoting a vector normal to the string boundary), and higher derivative terms organized in inverse powers of the string length $L$; see \cite{Aharony:2013ipa} for a systematic account. 

This theory is in very good agreement with lattice studies. In particular, both the quark-antiquark potential and the energy spectrum of excited states that follows show surprisingly good agreement \cite{LuscherWeisz2,Billo:2010ix,Brandt:2009tc,Athenodorou:2010cs,Athenodorou:2011rx}, correctly accounting for the hadron spectrum up to $1/L^5$ in 4D. See \cite{DubovskyGorbenko, Dubovsky1, Dubovsky2} for recent discussions of this approach and \cite{Aharony:2013ipa} for a recent review. Our discussion below also shares some qualitative aspects with light-front holography that was put forward in \cite{deTeramond:2008ht}.  It suffices for our purposes here to assume that there exists such an {\em exact} or {\em effective} string description given by a 2D conformal field theory. We will not need the precise form of this CFT.

\section{Rapidity scaling of entanglement entropy}

The setting plotted in Fig.~\ref{fig1} instructs us to trace over $x^+$ in the effective 2D CFT. We are thus interested in computing the entanglement entropy $S_A=-\Tr \rho_A \log \rho_A$ for the reduced density matrix $\rho_A$ associated with a null interval $A$ along $x^+$, in a 2D CFT. This can be achieved in a number of ways \cite{Calabrese:2009qy}: i) using the replica trick, i.e., from the partition function on a $n$-sheeted Riemann surface, or ii) from a correlation function of the so-called twist operators. Both methods yield the universal result for the entanglement entropy for an interval of width $\Delta x$ on a constant-time slice in the vacuum \cite{Holzhey:1994we,Vidal:2002rm,Calabrese:2004eu}, $S_A=\frac{c}{3}\log(\frac{\Delta x}{\epsilon})$, where $c$ is the central charge of the theory and $\epsilon$ is a UV cutoff. Using the relativistic invariance of the vacuum, one can
 immediately write down the corresponding expression for the entanglement entropy for an interval on a generic time slice. The result is given by \cite{Kusuki:2017jxh}
\be
S_A=\frac{c}{6}\log\left(\frac{\Delta x^2-\Delta t^2}{\epsilon^2}\right)=\frac{c}{6}\log\left(-\frac{\Delta x_+\Delta x_-}{\epsilon^2}\right)\,,
\ee
for an arbitrary spacelike interval with spatial and temporal widths $\Delta x$ and $\Delta t$, respectively. As shown in Fig. \ref{fig1}, we are interested in the null limit $\Delta x_-\to0$, where this expression has an IR divergence. This can be regulated by introducing a cutoff, $-\Delta x_-=\delta\ll\Delta x_+$, such that
\be
S_A=\frac{c}{6}\log\left(\frac{\delta \Delta x_+}{\epsilon^2}\right)\,.
\ee
The relation between light-cone and Milne coordinates is given by $x^\pm = \tau e^{\pm \eta}$, where $\tau$ and $\eta$ are proper time and rapidity, respectively. Since the proper time is fixed by the virtuality of the photon $Q^2$, $\tau\sim 1/\sqrt{Q^2}$, we get $\Delta x_+\sim\tau e^{\Delta\eta}$, and
in terms of these variables we obtain
\be\lab{SEE}
S_A=\frac{c}{6}\Delta\eta+\ldots\,,
\ee
where the ellipsis denotes terms that are independent of $\eta$. This is precisely the universal scaling behavior that was first derived in the parton model \cite{Kharzeev:2017qzs}. Here we see that it arises as a universal property of entanglement structure in 2D CFTs. 

On the other hand, we expect hadrons to correspond to excited states on the 2D CFT that are obtained by acting on the vacuum by vertex operators. We will now show that the same scaling law (\ref{SEE}) holds for these states, with a different prefactor. To start with, the reduced density matrix $\rho_A=\text{Tr}_{\bar{A}}\,\rho$ is Hermitian and positive semidefinite, so it can formally be expressed as
 \begin{equation} \label{modh}
\rho_A = \frac{e^{-H_A}}{\text{tr}\!\left( e^{-H_A} \right)} \ ,
 \end{equation}
where the Hermitian operator $H_A$ is known as the modular Hamiltonian. If we treat hadrons as perturbations of the vacuum, we have $\rho_A=\rho_A^{(0)}+\lambda\delta\rho_A$, and we can proceed as follows. To first order in the perturbation, we define the variation $\delta {\cal O}$ of any quantity $\mathcal{O}$ by $\delta {\cal O} = \partial_\lambda {\cal O}(\lambda)|_{\lambda = 0}$. In particular, the variation of entanglement entropy $S_A=-\text{Tr} \rho_A\log\rho_A$ is
given by $S_A=S_A^{(0)}+\lambda\delta S_A$, where
\begin{eqnarray}
\delta S_A &=& -\text{tr}\left[ \delta\rho_A \log \rho_A \right]-\text{tr} \left[\rho_A\, \rho_A^{-1}\delta\rho_A \right]\nonumber\\
\label{derfirst}
&=& \text{tr}\left[\delta\rho_A\, H_A \right]-\text{tr} \left[ \delta\rho_A \right]\,.
\end{eqnarray}
The last term in \eqref{derfirst} is identically zero, since the trace of the reduced density matrix equals 1 by definition, and it should remain normalized after introducing the perturbation. Hence, the leading order variation of the entanglement entropy yields
\begin{equation}
\label{firstlaw}
\delta S_A =\delta  \langle H_A \rangle\,,
\end{equation}
where $\langle H_A \rangle\equiv \text{tr}\left[\rho_A\, H_A \right]$. This is known as the first law of entanglement entropy \cite{HaagBook}. 

For a CFT, either free or interacting, the change in the modular Hamiltonian for the null segment $x_+\in[0,1]$ is given by \cite{Bousso:2014sda,Bousso:2014uxa}:
\be
\delta H_A=2\pi\int_0^1 dx_+ T_{++}(x_+)\,g(x_+)\,,
\ee
where $T_{++}(x_+)$ is the $++$ component of the energy-momentum tensor and $g(x_+)$ is a theory-dependent function that must satisfy a number of physical requirements. In particular, near each boundary of the segment, the entanglement structure should resemble that of Rindler space, for which $g(x_+) = x_+$, from the Bisognano-Wichmann theorem \cite{Bisognano:1976za}. Hence $g$ must satisfy i) $g(0) = 0$, ii) $g'(0) = 1$ and iii)
$g(1-x_+) = g(x_+)$. For the case of free CFTs in arbitrary dimensions, or generic 2D CFTs, the function $g$ is, in fact, uniquely determined and is given by \cite{Bousso:2014uxa}
\be
g(x_+)=x_+(1-x_+)\,.
\ee
We can also consider a segment $x_+\in[0,\Delta x_+]$. This amounts to the following transformation:
\be
\delta H_A=2\pi\Delta x_+\int_0^{\Delta x_+} dx_+ T_{++}(x_+)\,g(x_+/\Delta_{x_+})\,,
\ee
or, equivalently, if we denote $y\equiv x_+/\Delta x_+$
\be
\delta H_A=2\pi(\Delta x_+)^2\int_0^{1} dy\, T_{++}(y \Delta x_+)\,\tilde{g}(y)\,,\qquad \tilde{g}(y)=y(1-y)\,.
\ee
The first law (\ref{firstlaw}) then yields:
\be\label{deltaS}
\delta S_A=2\pi(\Delta x_+)^2\int_0^{1} dy\, \langle T_{++}(y \Delta x_+) \rangle \,y(1-y)\,.
\ee
We emphasize that this formula is valid when hadrons are treated as perturbative excitations of the vacuum. In general, the entanglement entropy can get contributions from other operators but the following inequality still holds \footnote{In higher dimensional CFTs, $\delta S_A = \delta  \langle H_A \rangle$ holds as an exact equality, even beyond linear order \cite{Bousso:2014uxa}.}:
\be
\delta S_A \leq \delta  \langle H_A \rangle\,.
\ee
This formula follows from the positivity of relative entropy.

To use the formula (\ref{deltaS}), we now need to compute $\langle T_{++}(x_+)\rangle$ for the state representing the hadron. The state-operator map in CFT allows for excited states to be expressed as vertex operators acting on the vacuum. We assume that these vertex operators are conformal primaries $V_h(x_+,x_-)$ \footnote{This assumption can be relaxed to {\em quasi-primaries} i.e. conformal fields invariant only under the SL(2,C) subgroup of conformal symmetry. Also, descendant states are implicitly covered as their correlation functions are fully determined from correlation functions of primaries.} with conformal weight $h$. Then the expectation value of the energy-momentum tensor in an excited state is determined by the three-point function $\langle V_{h}(y_+) T_{++}( x_+) V_{h}(z_+)\rangle$. The expectation value of the energy-momentum tensor in the excited state $\langle \phi_{\textrm{out}} | T_{++}(x_+) |\phi_{\textrm{in}} \rangle$ is then given by using the following relations between the in/out states and the vertex operators, see for example \cite{DiFrancesco:1997nk}: 
\be\lab{io}
|\phi_{\textrm{in}} \rangle = \lim_{z_+\to 0} V_{h}(z_+)|0\rangle\,, \qquad 
\langle \phi_{\textrm{out}} | = \lim_{y_+\to \infty} \langle 0 | V^\dagger_{h}(y_+) y_+^{2h}\, .
\ee
The reason for the extra power of $y_+$ in the out-state can be understood as arising from the conformal map from the cylinder, where the in- and out-states are defined naturally, to the complex plane. 
SL(2,C) invariance of the CFT fixes a generic three-point function as
$$\langle V_{h}(y_+) T_{++}( x_+) V_{h}(z_+)\rangle = \frac{C_{h}}{(y_+-x_+)^h(y_+-z_+)^{2h-2} (x_+-z_+)^h}\,,$$ 
where $C_h$ is the OPE coefficient. For example, for the operator $V = \exp(i\, k\cdot X(z))$ one has $h = k^2/2$ and $C_h = k^2/2$ where $k$ is transverse momentum in units of the square root of the string tension. To obtain this result we used the fact that the energy-momentum tensor is a quasiprimary operator with conformal weight $2$ and that momentum in the target space is conserved. Use of (\ref{io}) then immediately gives the desired result 
\be\lab{Tex} 
\langle \phi_{\textrm{out}} | T_{++}(x_+) |\phi_{\textrm{in}} \rangle = \frac{C_h}{x_+^2}\, .
\ee
However, if we use this profile in (\ref{deltaS}), the expression for $\delta S_A$ is formally divergent. The reason is that this energy-momentum tensor corresponds to an infinite energy excitation, $\int dx_+\langle T_{++}(x_+)\rangle\to\infty$, because the pointlike source introduces a nonintegrable singularity. We can regulate it by introducing a short-distance cutoff near the boundary of the entangling region $\delta$, which then yields
\be\label{deltaScft}
\delta S_A=2\pi C_h\left[\log \Delta x_+ -(1+\log\delta)+\mathcal{O}\left(\frac{1}{\Delta x_+}\right)\right],
\ee
or, in terms of rapidity,
\be
\delta S_A=2\pi C_h\Delta \eta +\cdots.
\ee
We thus recover the universal scaling of entanglement entropy imposed by conformal symmetry.

\subsection{States breaking conformal invariance\label{sec:states}} 
Alternatively, we can regulate the divergence of entanglement entropy by considering finite energy excitations that break conformal symmetry. In fact, it is enough to consider smearing out the vertex operator over a finite region, such that $\langle T_{++}(x_+)\rangle$ remains finite at $x_+\to0$ \footnote{A simple way to achieve this is to consider the state $e^{-\delta H}V_{h}|0\rangle$ where $H$ is the Hamiltonian and $\delta$ is a UV regulator that suppresses high energy modes \cite{Calabrese:2009qy}.}, i.e.,
\be
\langle T_{++}(x_+)\rangle\propto\left\{
                              \begin{array}{ll}
                                \frac{1}{x_+^2}, & x_+\to\infty \vspace{0.2cm}\\
                                \text{constant}, & x_+\to0
                              \end{array}
                            \right.
\ee
In this case, we arrive at the expected result, but we get subleading corrections that are nonuniversal and may reveal information about the hadron state. To give an example, consider a simple model where
\be
\langle T_{++}(x_+)\rangle=\frac{m}{\frac{m}{k}+x_+^2}=\left\{
                              \begin{array}{ll}
                                \frac{m}{x^2}, & x_+\to\infty \vspace{0.2cm}\\
                                k, & x_+\to0
                              \end{array}
                            \right.
\ee
Using (\ref{deltaS}) we then arrive at:
\bea
\delta S_A&\!\!=\!\!&2\pi m\left[\log\sqrt{1+\frac{k(\Delta x_+)^2}{m}}+\frac{1}{\Delta x_+}\sqrt{\frac{m}{k}}\arctan\left(\sqrt{\frac{k}{m}}\Delta x_+\right)-1\right]\\
&\!\!=\!\!&2\pi m\left[\log \Delta x_+ -\left(1+\log\sqrt{\frac{m}{k}}\right)+\mathcal{O}\left(\frac{1}{\Delta x_+}\right)\right],
\eea
from which we can recover the linear in rapidity scaling of entanglement entropy as the leading order. Further, we can identify $\delta=\sqrt{m/k}$ as the short distance cutoff introduced in (\ref{deltaScft}) or, equivalently, as the characteristic length scale over which the vertex operator is smeared out. Subleading corrections in rapidity appear at order $\mathcal{O}(e^{-\Delta \eta})$ and encode information about the particular state. In particular, they can be related on a one-to-one basis with the coefficients appearing in a long-distance expansion of the energy-momentum tensor.

\section{Discussion}
\lab{sec4} 

In this paper, we showed that the linear dependence of entanglement entropy on rapidity that was previously derived using the specific wave functions of hadrons based on the parton model \cite{Kharzeev:2017qzs}, arises naturally as a consequence of an approximate conformal invariance of the system inside the light cone. This is very much in line with the effective string description of hadron physics, which we shortly reviewed in section \ref{sec2}. Our result supports this picture and suggests that not only the traditional observables such as the quark-antiquark potential or the spectrum of excited states but also the entanglement structure inside hadrons can be captured by this simple string model. 

We emphasize that the leading term in the scaling law is independent of the details of the particular string theory, it only arises from the existence of an underlying 2D CFT description (exact or effective). In fact, we expect subleading corrections in $\eta$ to be nonuniversal and to contain crucial information about the specific string theory and the hadronic state. It will be useful to study these subleading terms and to contrast them with the ones that arise in the parton model to extend the entries of the duality dictionary. Our approach can also be refined by introducing deviations from conformal invariance. In general, such breaking of conformality would arise either by considering more general states, as exemplified in \ref{sec:states}, or by deforming the theory with a relevant operator. In the former case our formula relating the modular Hamiltonian to the entanglement entropy, (\ref{deltaS}) is valid, and our scaling law holds true as long as the energy-momentum tensor decays as $\propto 1/x_+^2$ at long distances. On the other hand, breaking conformal invariance by a relevant operator can be studied using conformal perturbation theory \cite{Speranza:2016jwt}, which we hope to turn to in the future. These calculations could similarly be carried out assuming an exact stringy description of hadrons, through the AdS/CFT correspondence. For example, in this context one could consider the entanglement entropy of a heavy meson propagating through a hot plasma \cite{Liu:2006nn,Chernicoff:2006hi}, which should be amenable to analytic computation for a variety of large-$N$ theories, e.g., \cite{Hubeny:2014zna}, and should admit a similar rapidity expansion, at least in the ultrarelativistic regime.

Finally, let us mention that our results could be used to provide an ``entropic" definition of the hadron structure function at small Bjorken $x$ that does not rely on weak coupling or on the validity of the parton model. Using the relation $S = \ln[xG(x)]$ between the entanglement entropy and the structure function \cite{Kharzeev:2017qzs}, we can define the latter as $xG(x) \equiv \exp[S]$. The linear dependence of the entanglement entropy then translates into the power growth of the structure function, $xG(x) \sim (1/x)^c$, where the constant $c$ is proportional to the central charge of the effective CFT.

\begin{acknowledgements}

UG is supported by the Netherlands Organisation for Scientific Research (NWO) under the VICI grant VI.C.202.104. 
The work of DK was supported by the U.S. Department of Energy, Office of Science, National Quantum Information Science Research Centers, Co-design Center for Quantum Advantage (C2QA) under Contract No.DE-SC0012704, and by the U.S. Department of Energy, Office of Science, Office of Nuclear Physics, Grants Nos. DE-FG88ER41450 and DE-SC0012704.
JFP is supported by the `Atracci\'on de Talento' program (Comunidad de Madrid) grant 2020-T1/TIC-20495, by the Spanish Research Agency via grants CEX2020-001007-S and PID2021-123017NB-I00, funded by MCIN/AEI/10.13039/501100011033, and by ERDF A way of making Europe. We are grateful to IFT UAM/CSIC and ECT* Trento for their hospitality, where the idea behind this paper was conceived and part of the project was carried out. 

\end{acknowledgements}

\addcontentsline{toc}{section}{References}

\bibliographystyle{apsrev4-2}
\bibliography{refs-hadrons}

\end{document}